\documentclass[iop,appendixfloats]{emulateapj}
\usepackage[backref,breaklinks,colorlinks,citecolor=blue]{hyperref} 
\usepackage[all]{hypcap} 

\usepackage{graphics}
\usepackage{graphicx}
\usepackage{rotating}
\usepackage{gensymb}
\usepackage{textcomp}
\newcommand{\extravspace}{\rule{0pt}{2.7ex}}

\newcommand{\um}{$\mu$m~}

\shorttitle{}
\shortauthors{Rich et al.}

\begin{document}

\title{\emph{The Carnegie-Chicago Hubble Program:} Calibration of the Near-Infrared RR Lyrae Period-Luminosity Relation With \emph{HST}}

\author{Jeffrey~A.~Rich}
\affiliation{The Observatories of the Carnegie Institution for Science, 813 Santa Barbara St., Pasadena, CA 91101, USA}
\author{Barry~F.~Madore}
\affiliation{The Observatories of the Carnegie Institution for Science, 813 Santa Barbara St., Pasadena, CA 91101, USA}
\author{Andrew~J.~Monson}
\affiliation{Department of Astronomy \& Astrophysics, The Pennsylvania State University, 525 Davey Lab, University Park, PA 16802, USA}
\author{Wendy~L.~Freedman}
\affiliation{Department of Astronomy \& Astrophysics, University of Chicago, 5640 South Ellis Avenue, Chicago, IL 60637, USA}
\author{Rachael L. Beaton}
\altaffiliation{Hubble Fellow}
\altaffiliation{Carnegie-Princeton Fellow}
\affiliation{Department of Astrophysical Sciences, Princeton University, 4 Ivy Lane, Princeton, NJ~08544}
\author{Gisella Clementini}
\affiliation{INAF-Osservatorio di Astrofisica e Scienza dello Spazio, via Piero Gobetti 93/3, I-40129, Bologna, Italy}
\author{Alessia Garofalo}
\affiliation{INAF-Osservatorio di Astrofisica e Scienza dello Spazio, via Piero Gobetti 93/2, I-40129, Bologna, Italy}
\affiliation{Dipartimento di Fisica e Astronomia, Universit\'a di Bologna, Viale Berti Pichat 6/2, I-40127 Bologna, Italy}
\author{Dylan Hatt}
\affiliation{Department of Astronomy \& Astrophysics, University of Chicago, 5640 South Ellis Avenue, Chicago, IL 60637, USA}
\author{Taylor Hoyt} 
\affiliation{Department of Astronomy \& Astrophysics, University of Chicago, 5640 South Ellis Avenue, Chicago, IL 60637, USA}
\author{In-Sung~Jang}
\affiliation{Department of Physics \& Astronomy, Seoul National University, Gwanak-gu, Seoul 151-742, Korea}
\author{Juna~A.~Kollmeier}
\affiliation{The Observatories of the Carnegie Institution for Science, 813 Santa Barbara St., Pasadena, CA 91101, USA}
\author{Myung~Gyoon~Lee}
\affiliation{Department of Physics \& Astronomy, Seoul National University, Gwanak-gu, Seoul 151-742, Korea}
\author{Jillian~R.~Neeley}
\affiliation{University of Florida, Department of Astronomy, 211 Bryant Space Science Center P. O. 112055, Gainesville, FL, USA}
\author{Victoria~Scowcroft}
\affiliation{Department of Physics, University of Bath, Claverton Down, Bath, BA2 7AY, UK}
\author{Mark~Seibert}
\affiliation{The Observatories of the Carnegie Institution for Science, 813 Santa Barbara St., Pasadena, CA 91101, USA}
 
\email{jrich@carnegiescience.edu}

\date{\today}

\pagebreak
\pagebreak

\begin{abstract}
We present photometry of 30 Galactic RR Lyrae variables taken with \emph{HST} WFC3/IR for the \emph{Carnegie-Chicago Hubble Program}. These measurements form the base of the distance ladder measurements that comprise a pure Population II base to a measurement of $H_{o}$ at an accuracy of 3\%. These data are taken with the same instrument and filter (F160W) as our observations of RR Lyrae stars in external galaxies so as to to minimize sources of systematic error in our calibration of the extragalactic distance scale. We calculate mean magnitudes based on one to three measurements for each RR Lyrae star using star-by-star templates generated from densely time-sampled data at optical and mid-infrared wavelengths. We use four RR Lyrae stars from our sample with well-measured \emph{HST} parallaxes to determine a zero point. This zero point will soon be improved with the large number of precise parallaxes to be provided by Gaia. We also provide preliminary calibration with the TGAS \& Gaia DR2 data, and all three zero points are in agreement, to within their uncertainties.
\end{abstract}


\section{Introduction}
Since the discovery of the expanding Universe by \citet{Hubble29} astronomers have measured the rate of expansion, the Hubble Constant ($H_{o}$), with increasing accuracy. The completion of the Hubble Space Telescope Key Project established the modern era of $H_{o}$ measurements, with more recent Cepheid-based distance ladder studies reaching the 3\% level of precision \citep{Freedman01,Freedman12,Riess16}. The \emph{Carnegie-Chicago Hubble Program} (CCHP) aims to provide a new and independent pathway to the measurement of $H_{o}$.  Using a distance ladder comprised of RR Lyrae stars, Tip of the Red Giant Branch (TRGB) stars and Type Ia Supernovae (SNe Ia), CCHP-II will provide a pure Population II basis to a measurement of $H_{o}$, with an aim of achieving an accuracy of 3\% (\citealt{Beaton16}, Freedman et al. 2018, in preparation). In this way, galaxies with only old stellar populations (and thus no Cepheid variables) can also be used as an independent assessment of the ongoing tension between distance-ladder based and Cosmic Microwave Background-based measurements of $H_{o}$ \citep{Freedman17,Riess18}. The SNe Ia zero-point can be set using only the TRGB rung of the distance ladder, with both current TRGB parallax measurements (e.g. \citealt{Brown18}) and in the future with \emph{Gaia} parallaxes. The TRGB zero-point itself, however, can be independently set by the RR Lyrae stars. Thus the first step in this distance ladder will be enabled by a well-calibrated, accurate zero point for Galactic RR Lyrae stars.

The existence of a near-infrared PL relation for RR Lyrae stars was first demonstrated by \citet{Longmore86,Longmore90} with K-band observations of RR Lyrae stars in Galactic globular clusters. Theoretical models now also clearly show a PL relation for RR Lyrae stars is to be expected at {nearly} all wavelengths, {with the exception of V-Band (e.g. \citealt{Bono03, Catelan04})}. Further, \citet{Madore12} demonstrated that there is a decrease in intrinsic scatter with an increase in wavelength of PL relations for both Cepheids and RR Lyrae stars due to a number of effects including a lower sensitivity of surface brightness to temperature variations and, to a lesser extent, a decrease in the line-of-sight extinction. As such, IR-wavelength determinations of the PL relation for RR Lyrae stars have significant benefits when compared to visible light measurements.

RR Lyrae PL relations are well characterized from optical (R \& I) to mid-IR wavelengths (e.g. \citealt{Madore13, Dambis14, Braga15, Neeley15}), but in order to use RR Lyrae stars as part of a precision distance ladder in the determination of $H_{o}$, it is essential to reduce both random and systematic errors \citep{Freedman12, Beaton16, Riess16}. As such, the CCHP was designed to employ the smallest number of instruments and filter combinations possible across the entire distance ladder. To minimize systematics the CCHP $H_{o}$ calibration, in fact, seeks a purely space-based calibration of $H_{o}$ with the use of a single telescope, the Hubble Space Telescope (\emph{HST}). This is accomplished by calibrating the RR Lyrae PL relation at the reddest near-IR wavelengths available to \emph{HST}, with the F160W (H-band) filter on the the WFC3/IR camera.

A near-IR calibration of the Galactic RR Lyrae PL zero point is now possible given the availability of accurate ($\sim$8\%) parallaxes determined for a few stars using the \emph{HST} Fine Guidance Sensor (FGS) (\citealt{Benedict11}) and in the near future \emph{Gaia} will provide parallaxes at similar or better precision for hundreds of RR Lyrae stars \citep{Clementini16,Gaia17}. In this paper we present our measurement of the geometrically-based period-luminosity relation for Galactic RR Lyrae stars with \emph{HST} WFC3/IR photometry that can now be used to calibrate this particular first rung of the CCHP-II distance ladder.

\section{Sample \& Observations}
Our sample consists of 30 Galactic RR Lyrae stars. Five of our RR Lyrae stars have trigonometric parallaxes measured with HST/FGS: XZ Cyg, UV Oct, RR Lyr, SU Dra and RZ Cep \citep{Benedict11}. The remaining 25 Galactic RR Lyrae stars will ultimately have their parallaxes determined by \emph{Gaia} with an error of $\sim$1\% \citep{deBruijne14,Clementini16,Gaia17}. Our observations were conducted with \emph{HST} WFC3/IR and were spread across two \emph{HST} programs as described in detail in the following subsections. The first program consisted of single-epoch SNAP observations designed to form the foundation of our full HST-based calibration (PID 13472, PI Freedman). The second program was designed to link the RR Lyrae stars and TRGB calibrations out to the Hubble flow, with dedicated follow-up and single-epoch observations for the five RR Lyrae stars with accurately measured parallaxes (PID 13691, PI Freedman).  The final sample is presented in Table 1, periods are adopted from \citet{Monson17}, wherein a self-consistent solution for multi-wavelength, long time-baseline data was found for each star in our sample. All of our stars have B, V, I, and {Spitzer} 3.6 \& 4.5 \um data.

\subsection{SNAP Observing Program}
The majority of our observations were taken in the course of our \emph{HST} SNAP program. Each star observed in the SNAP program has only one randomly timed epoch due to the nature of SNAP observations. With the exception of RR Lyrae itself, our data consist of several very short direct exposures due to the brightness of our targets and the desire to gather as much data as possible in the portion of each \emph{HST} orbit we were allotted.

\begin{table*}
\caption{Galactic RR Lyrae Sample \label{table_observations}}
\centering
\begin{tabular}{llllccc}
\multicolumn{1}{c}{Star}  & \multicolumn{1}{c}{Period (d)} & \multicolumn{1}{c}{Observations} & \multicolumn{1}{c}{Type}  & \multicolumn{1}{c}{Blazhko} & \multicolumn{1}{c}{[Fe/H]} & \multicolumn{1}{c}{A$_{F160W}$}\\
\hline\extravspace
SW And      & 0.442270  & a        &  RRab &  *    & -0.24  &  0.021 \\  
UY Cam      & 0.267042  & a        &  RRc  &       & -1.33  &  0.012 \\
RZ Cep      & 0.308710  & a,c,d    &  RRc  &       & -1.77  &  0.043 \\
RR Cet      & 0.553028  & a        &  RRab &       & -1.45  &  0.012 \\
XZ Cyg      & 0.466599  & a,c,d    &  RRab &  *    & -1.44  &  0.053 \\
DX Del      & 0.472617  & a        &  RRab &       & -0.40  &  0.051 \\
SU Dra      & 0.660420  & a,c,d    &  RRab &       & -1.80  &  0.006 \\
CS Eri      & 0.311331  & a        &  RRc  &       & -1.41  &  0.010 \\
SV Eri      & 0.713796  & a        &  RRab &       & -1.70  &  0.047 \\
RR Gem      & 0.397290  & a        &  RRab &  *    & -0.29  &  0.030 \\
TW Her      & 0.399600  & a        &  RRab &       & -0.69  &  0.023 \\
BX Leo      & 0.362860  & a        &  RRc  &       & -1.28  &  0.013 \\
RR Leo      & 0.452393  & a        &  RRab &       & -1.60  &  0.020 \\
TT Lyn      & 0.597434  & a        &  RRab &       & -1.56  &  0.009 \\
RR Lyr      & 0.566838  & a,b,c,d  &  RRab &  *    & -1.39  &  0.017 \\
UV Oct      & 0.542580  & a,c,d    &  RRab &  *    & -1.74  &  0.050 \\
BH Peg      & 0.640993  & a        &  RRab &  *    & -1.22  &  0.043 \\
RU Psc      & 0.390385  & a        &  RRc  &  *    & -1.75  &  0.024 \\
HK Pup      & 0.734238  & a        &  RRab &       & -1.11  &  0.088 \\
RU Scl      & 0.493339  & a        &  RRab &       & -1.27  &  0.010 \\
SV Scl      & 0.377340  & a        &  RRc  &       & -1.77  &  0.008 \\
AN Ser      & 0.522072  & a        &  RRab &       & -0.07  &  0.022 \\
V440 Sgr    & 0.477479  & a        &  RRab &       & -1.40  &  0.047 \\
V675 Sgr    & 0.642289  & a        &  RRab &       & -2.28  &  0.072 \\
MT Tel      & 0.316900  & a        &  RRc  &       & -1.85  &  0.021 \\
AM Tuc      & 0.405810  & a        &  RRc  &  *    & -1.49  &  0.013 \\
AB UMa      & 0.599577  & a        &  RRab &       & -0.49  &  0.012 \\
RV UMa      & 0.468060  & a        &  RRab &  *    & -1.20  &  0.010 \\
SX UMa      & 0.307118  & a        &  RRc  &       & -1.81  &  0.006 \\
TU UMa      & 0.557659  & a        &  RRab &       & -1.51  &  0.012 \\
      
\hline
\hline
\end{tabular}
\begin{quote}
  (a) Direct imaging in SNAP program \#13472\\
  (b) Drift-scan imaging in SNAP program \#13472\\ 
  (c) Direct imaging in GO program \#13691\\
  (d) Drift-scan imaging in GO program \#13691
\end{quote}
\end{table*}

For every star it was necessary to use both the subarray and RAPID sampling readout modes to avoid saturating the detector. Based on saturation times estimated from the WFC3/IR exposure time calculator (ETC), stars were observed with either a 64x64 or 128x128 subarray with a single RAPID sample, resulting in the shortest possible total exposure times of 0.061 and 0.113 seconds respectively.  These rapid observations are further facilitated by the lack of a shutter for the WFC3/IR channel. This was done 26 times in sequence with a simple 2-point dither pattern, achieving the maximum number of exposures possible before a read-out of the WFC3/IR buffer was necessary.  The very short exposure and subarray size ensured that in all frames the only object with any detectable flux was the RR Lyrae variable targeted.

Based on the previous observations of bright Cepheids (\citealt{Riess14} and \citealt{Casertano16}), we observed the star RR Lyr using WFC3/IR in a drift-scanning mode. In order to drift rapidly enough that RR Lyr did not saturate, our observations were carried out with a full-frame readout in gyro-only guiding mode, facilitating a scan rate of 7.5 arcsec s$^{-1}$. With the maximum number of RAPID samples possible as the star sped across the detector, our total time spent on target for a single drift scan of RR Lyrae was 17.6 seconds. This observation was repeated four times with a two-point dither pattern, with the number of exposures again limited by the maximum number of full-frame observations possible before a buffer read was required.

\subsection{GO Observing Program}
The remaining observations of our sample were taken as part of our large \emph{HST} General Observing program summarized in \citet{Beaton16}. We used WFC3/IR to observe the five Galactic RR Lyrae stars from \citep{Benedict11}. This was done to achieve two sets of measurements separated in phase in order to better constrain the mean magnitudes of our zero-point calibrators and to allow us to assess the accuracy of our single-phase SNAP-based magnitudes.

We found that despite the predictions of the ETC, even the shortest exposures in our SNAP observations with WFC3/IR show evidence of non-linearity. Because of this, our observations included drift-scanning observations to avoid reaching non-linearity in an effort to cross-check our short-exposure SNAP and GO observations. We split a single orbit between direct and drift scanning observations. One star, UV-Oct, was allocated two orbits as it had not yet been observed in our SNAP program, while the remaining four stars were allocated one orbit each.

As with the SNAP observations, we observed each of the five targets with the smallest 64x64 subarray and RAPID sampling to minimize the number of pixels subject to non-linearity, resulting in 0.061 second individual exposures. We used a four-point dither pattern for these observations with 8 cycles over a single orbit, resulting in 32 individual exposures of each star.

After the direct imaging exposures, we again used gyro-only guiding mode to observe in drift scanning mode with full frame readout. Each star was observed in this manner with four exposures on different portions of the chip, with a total exposure time of 14.66 seconds for each, with a scan rate of 7.5 arcsec s$^{-1}$ for UV Oct and 6.5 arcsec s$^{-1}$ for the remaining four stars.

\section{Analysis}
Although none of our direct imaging observations contain truly saturated pixels (with the exception of RR Lyr), some of our observations have pixels which surpass the non-linearity threshold established in the \emph{HST} WFC3/IR manual and are likely approaching saturation between the zeroth and first (and only) read, resulting in an estimation of counts from the zeroth read only and a significant uncertainty in the total number of counts \citep{Riess11}. For these frames the estimation of counts is significantly overestimated, and the non-linearity correction applied by the \emph{HST} pipeline to the calibrated, flat-fielded data results in significant deviations from expected values. That is, simple aperture photometry of observations with these significantly brighter pixels results in highly inaccurate measurements and suggests that the actual flux is over-predicted in such cases. 

\begin{figure}[htbp]
\centering
\includegraphics[width=0.45\textwidth]{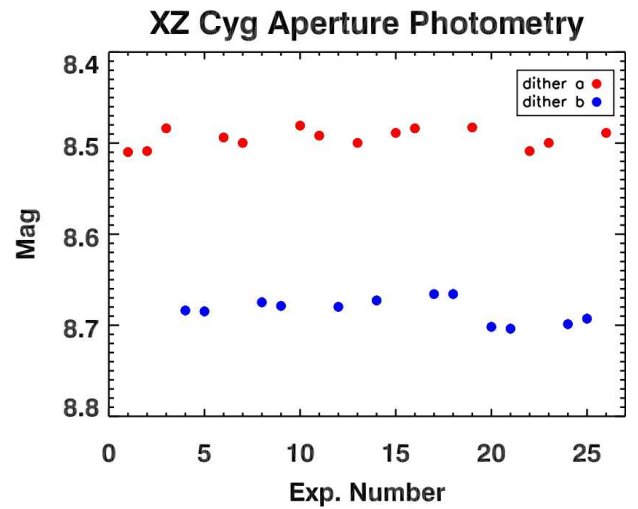}
\caption{Aperture photometry of the SNAP observations of XZ Cyg. Each point corresponds to a single frame, points are color-coded based on its dither position. The $\sim$0.2 mag difference caused by the dither pattern is a result of an overcorrection for non-linearity in a single bright pixel at dither point a.}
\end{figure}

Figure 1 shows each of the direct images of XZ Cyg obtained in the SNAP program; 12 of the exposures do not approach the ``saturation'' threshold (blue dots), while 14 of the exposures have one pixel with a significant non-linearity correction applied (red dots). Aperture photometry of the individual \emph{HST}-calibrated FLT files shows a a bi-modal pattern with a systematic difference of 0.2 mag. The bi-modality is a result of the dither pattern: in the exposures with a correction applied, the PSF of the star is centered perfectly on a single pixel while in the other exposures the center falls in a gap between pixels, lowering the peak flux in any one pixel below the non-linearity threshold.

Although the stars are well isolated, the presence of significantly non-linear pixels without a proper correction means that we cannot apply simple aperture photometry to our direct SNAP and GO observations. Our solution to this situation is to mask pixels in frames with poor non-linearity corrections and then undertake PSF photometry.

\subsection{Direct Imaging Photometry}
Photometry was carried out with DAOPHOT (\citealt{Stetson87, Stetson94}) on all individual direct imaging ``FLT'' frames with the \emph{HST} Pixel Area Map correction applied. Because we cannot generate a unique PSF for each frame, we used DAOPHOT to generate and apply one PSF for use in the 64x64 images and then generated a separate PSF for the 128x128 images. To generate the PSF, we used stars that are ``clean'' (i.e. images that do not have pixels that approach the non-linear regime): 156 images for the 64x64 subarray and 208 for the 128x128 subarray or 26 images of 6 and 8 stars each, respectively. 

To determine which parameters to assign to our model PSF, we generated several thousand individual PSF models with PSF radius and fitting radius varied in integer pixel increments from 2 to 17 and the 7 different PSF functional forms available in DAOPHOT. We then compared the magnitude derived from each model PSF for each star with aperture photometry and chose PSF parameters that, on the average, best reproduced the aperture magnitudes. We assigned a PSF radius of 14 pixels to the 64x64 PSF model and 17 pixels for the 128x128 PSF model and used a fitting radius of 12 pixels and a Moffat function with $\beta=3.5$ for both PSF models.

We also use the ``clean'' images to test the effect of pixel masking on our PSF photometry. We chose the brightest pixel in each image, masked it, and redetermined the magnitude using the same range of PSF parameter values. Regardless of the parameters used, masking pixels did not significantly impact DAOPHOT PSF photometry measurements. When masking stars in our 64x64 images we found a systematic offset of $-0.0017\pm0.0022$ mag, and for 128x128 images we found an offset of $-0.0024\pm0.0028$ mag.

Finally, we calculated aperture corrections to use with our PSF photometry with DAOGROW and aperture photometry of our ``clean'' images \citep{Stetson90}. We applied an aperture correction of 0.0196 mag to 64x64 images with a 14 pixel radius PSF and 0.0136 mag to 128x128 images with a 17 pixel PSF. These aperture corrections were computed to an infinite aperture, after which we applied the \emph{HST} WFC3/IR F160W Vega zero point for an infinite aperture of 24.695 mag (taken from \url{http://www.stsci.edu/hst/wfc3/ir_phot_zpt}). The final photometry for each observation, for each target, is given in Table 2. The errors quoted in Table 2 are the photometric errors derived by DAOPHOT. The mean magnitudes derived in Section 2 include an additional 2\% systematic error as described at \url{http://stsci.edu/hst/wfc3/phot_zp_lbn}.


\begin{table}[htpb!]
\caption{PSF photometry of individual exposures.} \label{table phot}
\centering
\begin{tabular}{lcccc}
\hline
\hline
\multicolumn{1}{l}{Name} & \multicolumn{1}{c}{MJD$_{obs}$} & \multicolumn{1}{c}{Phase} & \multicolumn{1}{c}{$F160W$} & \multicolumn{1}{c}{$F160W$}   \\
\multicolumn{1}{l}{} & \multicolumn{1}{c}{} & \multicolumn{1}{c}{} & \multicolumn{1}{c}{{mag}} & \multicolumn{1}{c}{$\sigma$}   \\
\hline
    AM Tuc   &     56900.43533022   &     0.98375   &    10.677   &    0.030  \\
    AM Tuc   &     56900.43553856   &     0.98426   &    10.676   &    0.030  \\
    AM Tuc   &     56900.43574688   &     0.98477   &    10.651   &    0.030  \\
    AM Tuc   &     56900.43642986   &     0.98646   &    10.699   &    0.030  \\

\hline
\hline
\end{tabular}
\\
{Example of photometry of individual exposures. Table is given in its entirety in machine readable format.}
\end{table}

\section{Mean Magnitude Determination}

\begin{table*}[htpb!]
\caption{Mean Magnitudes} \label{table phot}
\centering
\begin{tabular}{lccccc}
\hline
\hline
\multicolumn{1}{l}{Name} & \multicolumn{1}{c}{${F160W}$} & \multicolumn{1}{c}{${F160W}$} & \multicolumn{1}{c}{${F160W}$} & \multicolumn{1}{c}{${F160W}$} &  \multicolumn{1}{c}{${2MASS}$} \\
\multicolumn{1}{l}{} & \multicolumn{1}{c}{Spitzer} & \multicolumn{1}{c}{$\sigma_{s}$} & \multicolumn{1}{c}{Template} & \multicolumn{1}{c}{$\sigma_{T}$} &  \multicolumn{1}{c}{H} \\
\multicolumn{1}{l}{} & \multicolumn{1}{c}{mag} & \multicolumn{1}{c}{mag} & \multicolumn{1}{c}{mag} & \multicolumn{1}{c}{mag} &  \multicolumn{1}{c}{mag} \\

\hline
SW And     &   8.606 &   0.020 &   8.598 &   0.020 &   8.590 \\
UY Cam     &  10.851 &   0.024 &  10.849 &   0.025 &  10.841\\
RZ Cep     &   7.972 &   0.016 &   8.007 &   0.017 &   8.136 \\
RR Cet     &   8.609 &   0.025 &   8.617 &   0.023 &   8.652 \\
XZ Cyg     &   8.665 &   0.020 &   8.749 &   0.025 &   8.770 \\
DX Del     &   8.752 &   0.027 &   8.750 &   0.064 &   8.818 \\
SU Dra     &   8.701 &   0.018 &   8.733 &   0.019 &   8.686 \\
CS Eri     &   8.174 &   0.024 &   8.217 &   0.025 &   8.175 \\
SV Eri     &   8.657 &   0.023 &   8.718 &   0.024 &   8.736 \\
RR Gem     &  10.334 &   0.023 &  10.339 &   0.021 &  10.271 \\
TW Her     &  10.385 &   0.041 &  10.352 &   0.027 &  10.269 \\
BX Leo     &  10.688 &   0.034 &  10.750 &   0.058 &  10.743 \\
RR Leo     &   9.672 &   0.021 &   9.798 &   0.028 &   9.737 \\
TT Lyn     &   8.697 &   0.020 &   8.701 &   0.022 &   8.656 \\
RR Lyr     &   6.472 &   0.035 &   6.481 &   0.031 &   6.598 \\
UV Oct     &   8.316 &   0.013 &   8.292 &   0.013 &   8.289 \\
BH Peg     &   9.180 &   0.025 &   9.184 &   0.029 &   9.070 \\
RU Psc     &   9.190 &   0.025 &   9.244 &   0.021 &   9.156 \\
HK Pup     &  10.055 &   0.034 &  10.038 &   0.025 &  10.031 \\
RU Scl     &   9.252 &   0.025 &   9.286 &   0.020 &   9.285 \\
SV Scl     &  10.567 &   0.024 &  10.605 &   0.075 &  10.599 \\
AN Ser     &   9.895 &   0.026 &   9.895 &   0.020 &   9.914 \\
V440 Sgr   &   9.104 &   0.025 &   9.218 &   0.031 &   9.230 \\
V675 Sgr   &   9.088 &   0.025 &   9.114 &   0.022 &   9.059 \\
MT Tel     &   8.055 &   0.055 &   8.135 &   0.034 &   8.193 \\
AM Tuc     &  10.677 &   0.028 &  10.713 &   0.029 &  10.704 \\
AB UMa     &   9.707 &   0.020 &   9.694 &   0.024 &   9.862 \\
RV UMa     &   9.889 &   0.043 &   9.911 &   0.034 &   9.862 \\
SX UMa     &  10.152 &   0.028 &  10.200 &   0.021 &  10.150 \\
TU UMa     &   8.735 &   0.020 &   8.747 &   0.021 &   8.717 \\

\hline
\hline
\end{tabular}
\end{table*}

Because we have only sparsely-sampled light curves, we have undertaken template fitting to determine mean magnitudes for our sample of RR Lyrae stars. Fortunately, the stars in our sample have a wealth of ground and space-based observations available for the construction of light curve templates constructed on a star-by-star basis  \citep{Monson17}. We consider two approaches: (1) we compare our observations to smooth light curves constructed from \emph{Spitzer} $3.6\mu$m observations and (2) we use predicted templates generated from optical V and I band ground-based observations, as first proposed and demonstrated for Cepheids in \citet{Freedman10a}. In each case, we averaged all of the data points available in each epoch of SNAP or GO observation into a single-epoch, intensity-averaged point, resulting in one to three \emph{HST} F160W points to which a template light curve can be compared. Phases for the data points used in mean magnitude determination are calculated using the periods and phase correction information and methodology to be outlined in Beaton et al. (in prep).

The smoothed \emph{Spitzer} light curves were generated by applying GLOESS (Gaussian-windowed LOcal regrESSion method, \citealt{Persson04,Monson17}) to 20-40 evenly spaced data points, resulting in a smoothly varying curve \citep{Monson17}. The mid-IR data are well suited for comparison to our near-IR observations, as we expect by the H-band that the light-curve shape is dominated by radius effects. As such, we simply offset in magnitude the 3.6\um GLOESS curve to fit the available \emph{HST} data points and take the mean of the offset light curve to be our F160W mean magnitude. Figure 2 demonstrates this process for UV Oct, the star in our sample with the most data points. The error on our mean magnitude measurement combines the random photometric error of our data, the error in the template, and the error on the phase estimate and its location with respect to the template. There is an additional possible systematic error induced by the difference in amplitude between 3.6\um and H-band. Measurements by \citet{Monson17} show the H-band has a larger amplitude with a median of $\sim0.03$ magnitudes, leading to a potential offset in magnitude of $\sim0.01$ magnitudes, depending on the phase of our measurements.

For comparison with our \emph{Spitzer}-derived magnitudes, we follow the same procedure with custom-generated template light curves for each RR Lyrae variable. We use near-infrared templates generated from V and I band data using a process described in \citet{Beaton16, Monson17} and Beaton et al. (in prep). The templates require well-sampled B, V \& I light curves that can be combined to trace the change in temperature and radius. Temperature-independent curves that predict the shape of the IR light curves are generated using the visible light. These curves predict the shape of the IR light curves which primarily trace radius, with a scaling in amplitude \citep{Freedman10a, Freedman10b}.

We offset the template to match the available data and measure the mean of the shifted ground-based template. The intensity-averaged measured mean magnitudes for both methods are given in Table 3. The two methods provide consistent mean magnitudes, with an average difference of 0.005$\pm$0.019 mag (mag$_{Sptz}$-mag$_{tmpl}$). Further, our F160W magnitudes are offset by 0.015$\pm0.06$ mag from the Two Micron All-Sky Survey (2MASS) H-band magnitudes, which is roughly consistent with the 0.0215$\pm0.0054$ mag difference calculated by \citet{Riess11}. 

\begin{figure}[htbp!]
\centering
\includegraphics[width=0.5\textwidth]{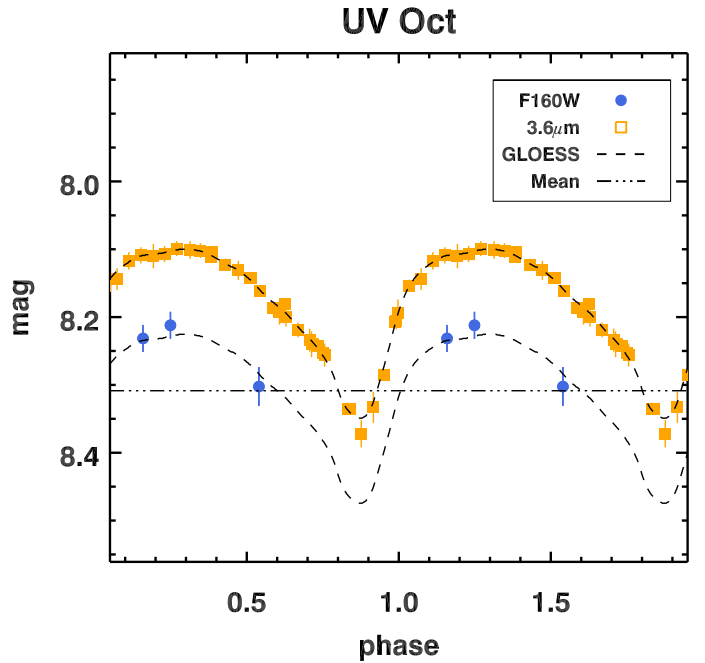}
\caption{Example of our \emph{HST} PSF photometry (blue points) for one RR Lyrae variable (UV Oct) compared to \emph{Spitzer} data and the corresponding GLOESS light curve (yellow squares and line). The plot shows the GLOESS fit to the Spitzer data prior to scaling (above) and after scaling (below) to the F160W data. The mean magnitude derived from the shifted curve is shown as a horizontal line.}
\end{figure}

\begin{figure*}[htbp!]
\includegraphics[width=0.50\textwidth]{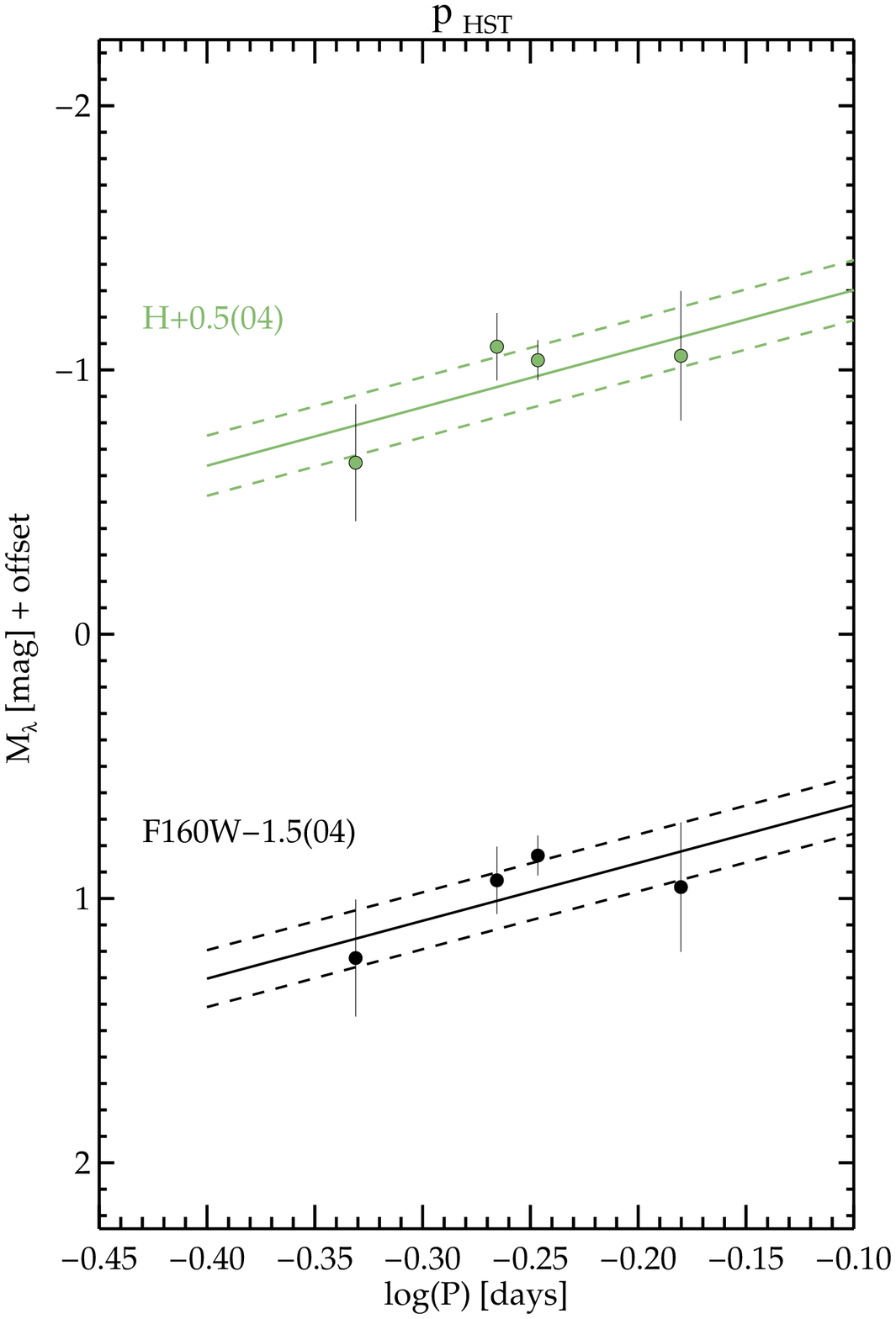}
\includegraphics[width=0.50\textwidth]{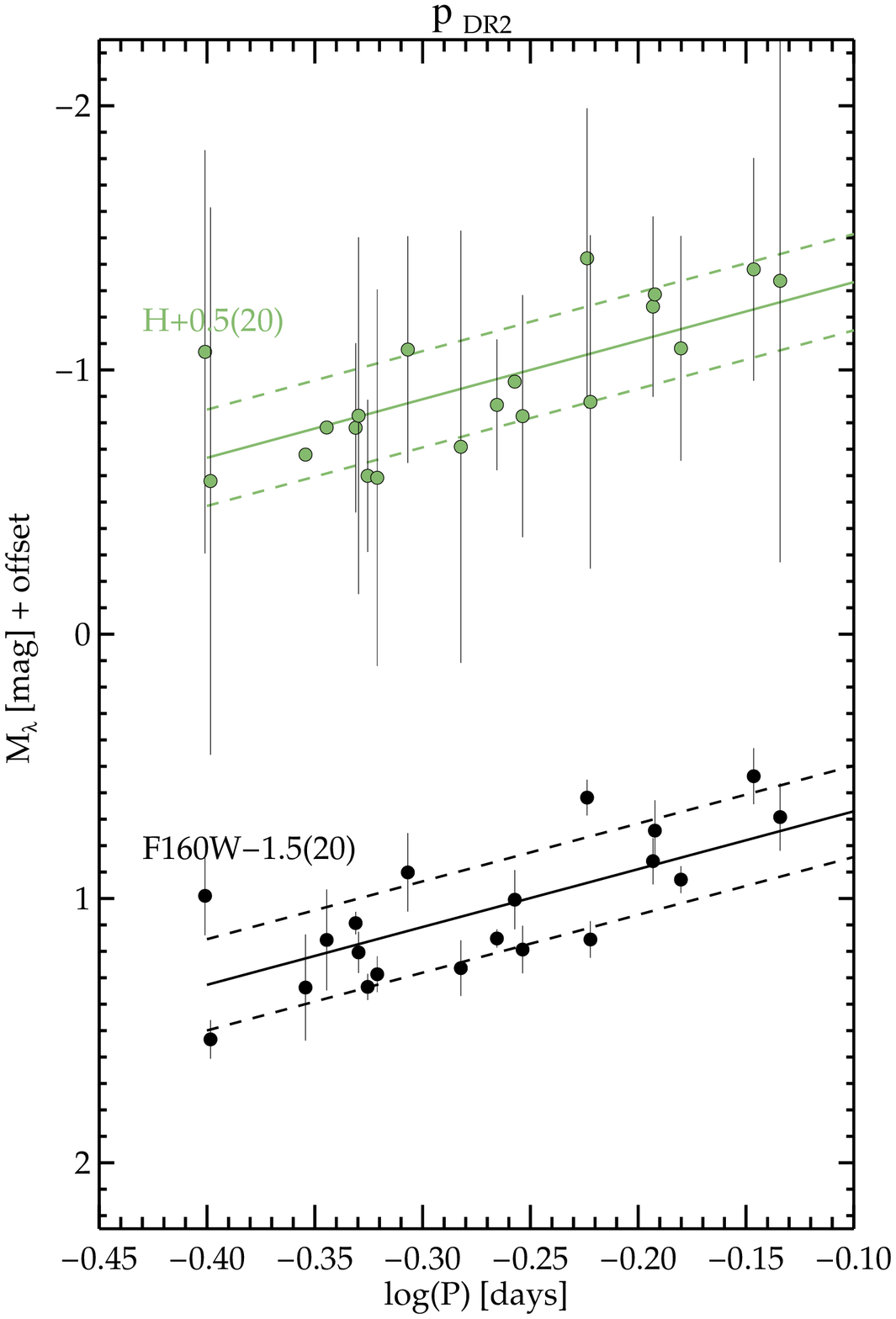}
\caption{Period-luminosity (PL) relations for our Galactic RR-Lyrae; only RRab (fundamental pulsators) are used. The left panel consists of the four RRab with \emph{HST}/FGS parallaxes from \citet{Benedict11}, { the right panel shows RRab in the \emph{Gaia} DR2 parallax sample, our F160W sample contains 20 stars (See section 5 for a discussion of our selection method)}. We compare our F160W PL relation with a fit to ground-based 2MASS H-band magnitudes \citep{Monson17}.}
\end{figure*}

\section{Period-Luminosity Relations}
To construct our PL relations we calculate absolute magnitudes for RR Lyrae stars from our F160W sample with known parallaxes. For this calculation we use the mean magnitudes derived by scaling \emph{Spitzer} data as described in the previous section. We examine RR Lyrae stars with \emph{HST}/FGS parallaxes (\citealt{Benedict11}), Tycho-\emph{Gaia} Astrometric Solution (TGAS) parallaxes (\citealt{Lindegren16,Gould17,Gaia17}) and \emph{Gaia} data release 2 (DR2) parallaxes \citep{gaia18,Lindegren18}. Our absolute magnitudes include a correction for extinction adopted from the published values of \citep{Feast08} that will be discussed further in Beaton et al. (in prep). The values we use in this paper are given in Table 1. As a check for consistency, we compare our absolute magnitudes and PL relations with 2MASS H-band magnitudes. We consider only the fundamental pulsators (RRab) in our PL analysis.

To compute the absolute magnitude, { we use two methods: we invert the parallax to produce a distance and we use the astrometry-based luminosity (ABL) method for parallaxes with larger errors. For direct inversion of parallaxes, we restrict our sample to $\sigma_{\pi}/\pi$ \textless 0.1, the regime where this is an appropriate treatment for the resulting distance uncertainties and no additional forward modeling is explicitly required (\citealt{Astraatmadja16,Gaia17}). We use only RRab stars for determination of the zero points \footnote{Although we do not consider RRc stars here, the multi-band analysis by \citet{Monson17} find an RRc zero point that is consistent with \citet{Kollmeier13}. These cuts result in a final sample of with 4, 3, and 20 RRab stars for \emph{HST}, TGAS, and DR2 parallaxes respectively.}}.

\begin{table}[htpb!]
\caption{Absolute Magnitudes} \label{table phot}
\centering
\begin{tabular}{lcccc}
\hline
\hline
\multicolumn{1}{l}{Name} & \multicolumn{1}{c}{H(mag)} & \multicolumn{1}{c}{[F160W]} &  \multicolumn{1}{c}{$\sigma$}  & Blazhko\\
\multicolumn{1}{l}{} & \multicolumn{1}{c}{${2MASS}$} & \multicolumn{1}{c}{${\emph{HST}}$} & & \multicolumn{1}{c}{mag}  \\
\hline
        & \textbf{\emph{HST}/FGS} & & &\\
XZ Cyg  & -0.149 & -0.275 &  0.221 & * \\
SU Dra  & -0.554 & -0.543 &  0.245 &   \\
RR Lyr  & -0.537 & -0.663 &  0.076 & * \\
UV Oct  & -0.588 & -0.569 &  0.127 & * \\
\hline
\hline
        & \textbf{TGAS} & &        &    \\
XZ Cyg  & -0.297 & -0.423 &  0.320 &  * \\
DX Del  & -0.132 & -0.198 &  0.288 &    \\
SU Dra  & -0.538 & -0.528 &  0.425 &    \\
SV Eri  & -0.157 & -0.164 &  0.421 &    \\
RR Gem  & -0.672 & -0.754 &  0.763 &  * \\
TT Lyn  & -0.226 & -0.207 &  0.568 &    \\
UV Oct  & -0.231 & -0.132 &  0.247 &  * \\
BH Peg  & -0.055 & -0.042 &  0.341 &  * \\
RU Scl  &  0.173 &  0.152 &  0.429 &    \\
RV UMa  & -0.806 & -0.775 &  0.675 &  * \\
TU UMa  & -0.459 & -0.440 &  0.458 &    \\
\hline
\hline
	& \textbf{Gaia DR2} & &          &   \\
SW And  & -0.180 & -0.163 &  0.200  & * \\
RR Cet  & -0.456 & -0.495 &  0.112  &   \\
XZ Cyg  & -0.281 & -0.407 &  0.043  & * \\
DX Del  & -0.099 & -0.166 &  0.050  &   \\
SU Dra  & -0.582 & -0.572 &  0.051  &   \\
SV Eri  & -0.881 & -0.963 &  0.106  &   \\
RR Gem  & -0.569 & -0.510 &  0.149  & * \\
TW Her  & -0.080 &  0.033 &  0.073  &   \\
RR Leo  & -0.282 & -0.343 &  0.191  &   \\
TT Lyn  & -0.923 & -0.882 &  0.068  &   \\
UV Oct  & -0.368 & -0.349 &  0.034  & * \\
BH Peg  & -0.740 & -0.641 &  0.088  & * \\
HK Pup  & -0.838 & -0.808 &  0.128  &   \\
RU Scl  & -0.577 & -0.599 &  0.148  &   \\
AN Ser  & -0.209 & -0.236 &  0.105  &   \\
V440 Sgr& -0.092 & -0.213 &  0.068  &   \\
V675 Sgr& -0.786 & -0.757 &  0.115  &   \\
AB UMa  & -0.379 & -0.345 &  0.070  &   \\
RV UMa  & -0.327 & -0.296 &  0.078  & * \\
TU UMa  & -0.325 & -0.307 &  0.090  &   \\
\hline
\hline
\end{tabular}
\end{table}

Absolute magnitudes { derived from directly inverted parallaxes} used in our PL analysis are given in Table 4. The errors for absolute magnitudes derived with \emph{HST}/FGS and TGAS parallaxes are calculated directly from the errors in the parallax as they dominate the uncertainty. The error in absolute magnitude when derived from Gaia DR2 parallaxes are a combination of the error on the mean magnitude (Table 3) and the error in parallax. The period-luminosity relation takes the simple form:

\begin{equation}
M_{\lambda} = a_{\lambda} \times [log(P) - 0.3] + b_{\lambda}
\end{equation}

\noindent where a and b are the slope and intercept respectively, and P is the period in days. We perform an unweighted least-squares (LSQ) fit assuming a fixed slope of a $ = $ $-$2.215 and a $ = $ $-$2.189 for fundamental pulsators in the 2MASS H-band (\citealt{Braga15}) and \emph{HST} F160W respectively (Marconi et al. in prep). Using the cuts discussed in the previous section four stars with \emph{HST}/FGS parallaxes, three with TGAS parallaxes, and 20 stars with DR2 parallaxes are used to generate different PL relations this way. The resulting zero points at log(P) = -0.3 days are given in Table 5.

\begin{table*}[htpb!]
\caption{Zero Points} \label{table phot}
\centering
\begin{tabular}{lcccc}
\hline
\hline
\multicolumn{1}{l}{Sample} & \multicolumn{1}{c}{Filter} &   \multicolumn{1}{c}{Method} & \multicolumn{1}{c}{Zero-point} & \multicolumn{1}{c}{Zero-point, No Blazhko}     \\
\hline
\emph{HST}/FGS & H     & LSQ         & $-0.36\pm0.07$   & $-0.29\pm0.25$ \\
\emph{HST}/FGS & F160W & LSQ         & $-0.42\pm0.06$   & $-0.28\pm0.25$ \\
\emph{HST}/FGS & F160W & LSQ, [Fe/H] & $-0.42\pm0.07$   & $-0.24\pm0.25$ \\
TGAS           & H     & LSQ         & $-0.30\pm0.08$   & ...            \\
TGAS           & F160W & LSQ         & $-0.30\pm0.09$   & ...            \\
TGAS           & F160W & LSQ, [Fe/H] & $-0.37\pm0.13$   & ...            \\
DR2            & H     & LSQ         & $-0.39\pm0.04$   & $-0.37\pm0.05$ \\
DR2            & F160W & LSQ         & $-0.39\pm0.04$   & $-0.38\pm0.05$ \\
DR2            & F160W & LSQ, [Fe/H] & $-0.47\pm0.04$   & $-0.44\pm0.04$ \\
\emph{HST}/FGS & H     & ABL         & $-0.40\pm0.14$   & $-0.43\pm0.29$ \\
\emph{HST}/FGS & F160W & ABL         & $-0.39\pm0.14$   & $-0.43\pm0.29$ \\
\emph{HST}/FGS & F160W & ABL, [Fe/H] & $-0.39\pm0.14$   & $-0.44\pm0.29$ \\
TGAS           & H     & ABL         & $-0.39\pm0.08$   & $-0.43\pm0.12$ \\
TGAS           & F160W & ABL         & $-0.39\pm0.08$   & $-0.43\pm0.12$ \\
TGAS           & F160W & ABL, [Fe/H] & $-0.38\pm0.08$   & $-0.42\pm0.12$ \\
DR2            & H     & ABL         & $-0.39\pm0.07$   & $-0.39\pm0.08$ \\
DR2            & F160W & ABL         & $-0.39\pm0.07$   & $-0.40\pm0.08$ \\
DR2            & F160W & ABL, [Fe/H] & $-0.37\pm0.07$   & $-0.38\pm0.08$ \\
\hline
\hline
\end{tabular}
\end{table*}

Example PL relation fits are shown in Figure 3. 

\subsection{Period-Luminosity-Metallicity Relations}
In order to examine the effect of metallicity on our zero point measurement we also consider a period-luminosity-metallicity (PLZ) relation for F160W. Our PLZ relation takes the following form:

\begin{equation}
M_{\lambda} = a_{\lambda} \times [log(P) - 0.3] + b_{\lambda} + c_{\lambda} \times ( [Fe/H] + 1.60 )
\end{equation}

\noindent where a, b and P take the same form as our PL relation, [Fe/H] is the metallicity of each RR Lyrae star (as compiled and homogenized by \citealt{Monson17}, from \citealt{Feast08}, \citealt{Fernley98}, and \citealt{Fernley97}, see Table 1.), and c is the slope of the adopted metallicity correction applied to each star, with a value of c$=$0.186 \citep{Braga15}. The results are shown in Table 5 for comparison. For the stars with \emph{HST} parallaxes, we find a fairly consistent zero point. RR Lyrae stars from the TGAS parallax sample exhibit a larger zero point discrepancy, but at present the errors on the absolute magnitudes result in too large a scatter to draw a significant conclusion about this difference.

\subsection{Astrometry-Based Luminosity}
In their analysis of RR Lyrae TGAS parallaxes, \citet{Gaia17} note that parallax inversion to calculate absolute magnitudes is inadvisable if the parallax errors are too large. We mirror their analysis here for comparison and include a measurement of the zero point using Astrometry-Based Luminosity (ABL, \citealt{Arenou99}), defined as:

\begin{equation}
a = 10^{0.2M} = \varpi^{0.2m_{o}-2}
\end{equation}

This is the same as \citet{Gaia17} formula (2), where $M$ is absolute magnitude, $\varpi$ is the parallax, $m_{o}$ is the extinction-corrected apparent magnitude. We then solve the following PL and PLZ relations for the zero point $b_{\lambda}$, corresponding to \citet{Gaia17} formulas (4) and (5):

\begin{equation}
10^{0.2(a_{\lambda}+b_{\lambda})} = \varpi10^{0.2m_{o}-2}
\end{equation}
\begin{equation}
10^{0.2(a_{\lambda}+b_{\lambda}+c_{\lambda}[Fe/H])} = \varpi10^{0.2m_{o}-2}
\end{equation}

We apply the same fixed slopes a and c used in sections 5.1 and 5.2, the resulting zero points are given in Table 5. {We use a less conservative cut of  $\sigma_{\pi}/\pi$ \textless 0.2. With this less stringent cut our usable TGAS sample includes 13 RRab stars for the ABL fit, while our \emph{HST} and DR2 samples remain the same size.}

\subsection{Blazhko Effect}
Several of the stars in our sample exhibit the Blazhko effect, which can affect the measurement of mean magnitudes in the optical as well as the IR (e.g. \citealt{Jurcsik18}). In order to consider the impact of the Blazhko pulsators sample, we fit PL relations with only those RRab stars which apparently do not suffer from amplitude modulation. {The fits are performed in the same manner as described above, with 1 and 14 stars from the \emph{HST} and Gaia DR2 parallax samples respectively. There are no non-Blazhko stars with TGAS parallaxes below the $\sigma_{\pi}/\pi$ \textless 0.10 cut, and 6 below the $\sigma_{\pi}/\pi$ \textless 0.20 cut. As such, we only calculate a zero point using the ABL method in this case.} The results are given in Table 5. The largest difference is for the \emph{HST} parallax sample when performing a least-squares fit, as the zero-point is now constrained by a single star. The ABL method shows consistent results between the zero points fit with and without Blazhko RRab.

\section{Conclusions}
We have carried out photometry using \emph{HST}/WFC3 of 30 Galactic RR Lyrae stars taken in F160W. Our results are part of the larger effort associated with the CCHP; the photometry presented here will be used to provide a calibration of the Galactic RR Lyrae zero point in the same flight-magnitude system used to measure RR Lyrae distances to nearby galaxies. The mean magnitudes that we calculate are consistent with ground-based H-band photometry, within the expected offset between the 2MASS and \emph{HST} flight magnitude systems.

We also present PL relations for the RRab stars in our sample using both \emph{HST}/FGS and Gaia TGAS \& DR2 parallaxes. These PL relations provide an initial estimate of the HST F160W zero point, with all three parallax samples providing zero points consistent within their respective errors. Our final Galactic RR Lyrae zero point will rely on a larger sample of more accurate parallaxes provided by the next \emph{Gaia} data release.

\section{Acknowledgements}
We thank the referee for their helpful and constructive comments. Myung Gyoon Lee was supported by the National Research Foundation of Korea (NRF) grant funded by the Korea Government (MSIP) (No. 2017R1A2B4004632). {This paper makes use of data from the European Space Agency (ESA) mission Gaia (https://www.cosmos.esa.int/gaia), processed by the Gaia Data Processing and Analysis Consortium (DPAC, https://www.cosmos.esa.int/web/gaia/dpac/consortium). Funding for the DPAC has been provided by national institutions, in particular the institutions participating in the  Gaia Multilateral Agreement.}

\bibliographystyle{apj}
\bibliography{ms}

\end{document}